\documentclass[conference]{IEEEtran}

\usepackage{cite}
\usepackage{psfrag}
\usepackage{url}
\usepackage{stfloats}
\usepackage{amsmath}
\usepackage{array}
\usepackage{epsfig}
\usepackage{amssymb}
\usepackage{color}
\usepackage{subfig}
\usepackage{setspace}
\usepackage{caption}
\usepackage{booktabs}
\usepackage{ dsfont }
\usepackage{algpseudocode}

\usepackage[square,numbers,sort&compress]{natbib}
\usepackage{algorithm}  
\usepackage{amsmath,epsfig,amssymb,algpseudocode,amsthm,cite,url}
\usepackage{comment}

\usepackage{multirow}
\DeclareMathOperator*{\argmax}{argmax}

\ifCLASSINFOpdf
\graphicspath{{"./pictures/"}}
\else
 \usepackage{enumerate}
 \usepackage{amsfonts,amssymb}
\fi

\begin{document}
\title{Efficient Drone Mobility Support Using Reinforcement Learning}

\author{\IEEEauthorblockN{Yun Chen$^1$, Xingqin Lin$^2$, Talha Khan$^2$, and Mohammad Mozaffari$^2$}
\IEEEauthorblockA{ \\
	$^1$The University of Texas at Austin, USA, Email: yunchen@utexas.edu \vspace{0.1cm}\\
	$^2$Ericsson Research, Santa Clara, CA, USA,  \\Emails: \{xingqin.lin, talha.khan, mohammad.mozaffari\}@ericsson.com
}
}

\maketitle 

\begin{abstract}
Flying drones can be used in a wide range of applications and services from surveillance to package delivery. To ensure robust control and safety of drone operations, cellular networks need to provide reliable wireless connectivity to drone user equipments (UEs). To date, existing mobile networks have been primarily designed and optimized for serving ground UEs, thus making the mobility support in the sky challenging. In this paper, a novel handover (HO) mechanism is developed for a cellular-connected drone system to ensure robust wireless connectivity and mobility support for drone-UEs. By leveraging tools from reinforcement learning, HO decisions are dynamically optimized using a Q-learning algorithm to provide an efficient mobility support in the sky. The results show that the proposed approach can significantly reduce (e.g., by 80\%) the number of HOs, while maintaining connectivity, compared to the baseline HO scheme in which the drone always connects to the strongest cell.

\end{abstract}

%
\IEEEpeerreviewmaketitle

\section{Introduction}
Owing to their mobility, agility, and flexibility, drones are widely used in various applications. In particular, drone user equipments (UEs) play a key role in a number of scenarios such as package delivery, remote sensing, and surveillance applications \cite{TutorialMO, fotouhi2019survey}.  For sustainable operation, flying drones need to be supported via cellular infrastructure (a.k.a. cellular-connected drones) to ensure seamless connectivity and low-latency communications. Cellular technologies such as Long-term Evolution (LTE) and the fifth-generation New Radio (5G NR) offer wide-area, high-speed, and secure wireless connectivity \cite{yang2018telecom, Sky}, which can provide robust control and safety for drone operations. %
In this regard,  there are several challenges in supporting drone-UEs in a cellular network. First,  drones can move in three-dimensions (3D), and their arbitrary trajectory and high speed result in rapid changes in received signal strength. Second, due to line-of-sight (LOS) propagation conditions, drones may suffer from strong uplink and downlink interference from neighbor cells \cite{TutorialMO}.
Third, terrestrial base stations (BSs) are mainly designed to serve ground users and hence their antennas are down-tilted. The main lobe of a BS antenna thus covers a large part of the surface area of the cell to improve performance for terrestrial UEs. Accordingly, at ground level the strongest site is typically the closest one. A drone-UE on the other hand may be frequently served by the sidelobes of BS antennas \cite{MobilieDrones}, which have s low antenna gain compared to the main lobe of the antennas. The coverage areas of the sidelobes may be small and the signals at the edges may drop sharply due to deep antenna nulls. At a given location, the strongest signal might come from a faraway BS, if  the sidelobes of the  BSs closer to the drone-UE is significantly weaker. Additionally, the sidelobes of BSs may not fully cover the sky and there can be coverage holes (space without coverage service) in the sky that can cause connectivity  failure. Meanwhile, the fragmented coverage area provided by different BSs hardens the mobility support in the sky and can result in frequent handovers (HOs). This, in turn, leads to significant signaling overhead and radio link failure (RLF) due to undesired ping-pong HOs.  Therefore, there is a need for efficient HO mechanisms for drone mobility management to provide reliable communications between drones and BSs.

\subsection {Related Work}

 In the Third-generation partnership project (3GPP) Release 15, the potential support of LTE for providing drone connectivity was studied \cite{3GPPTR}. The results of this study showed that mobility support for drones is one of the challenging aspects  in using existing LTE networks to serve drone-UEs. The work in \cite{stanczak2018mobility} identified key challenges associated with supporting drone connectivity in LTE networks. In \cite{MobilityS}, the performance of a cellular-connected drone network was evaluated in terms of RLF and the number of HOs.  In \cite{yajnanarayana20195g},   a handover optimization scheme was proposed for ground UEs in a 5G cellular network using reinforcement learning (RL). In \cite{alkhateeb2018machine}, the authors proposed a handover mechanism based on deep learning to improve the reliability and latency in terrestrial  millimeter-wave mobile  systems. While previous work has studied various other challenges related to drone communications, the problem of  handover optimization for drone-UEs in the sky remains an open problem. 

\subsection{Contributions}
In this paper, we propose a novel HO optimization mechanism for a cellular-connected drone system to ensure robust wireless connectivity for drone-UEs. By using tools from RL \cite{sutton1998introduction}, HO decisions are dynamically  optimized using Quality-learning (Q-learning) to provide an efficient mobility support in the sky. The proposed framework 
leverages the reference signal received power (RSRP) data and the drone's trajectory information 
to provide effective HO rules for seamless drone connectivity while considering HO signaling overhead. Furthermore, our results depict the inherent
trade-off between the number of HOs and the serving cell RSRP in the considered cellular-connected drone system.

The rest of this paper is organized as follows. Section II presents the system model. A brief background of RL relevant to our model is introduced in Section \ref{Background of RL}. In Section \ref{Model and Algorithms}, we present our RL-based HO scheme. The simulation results are provided in Section \ref{Experiments and Results}. Section \ref{Conclusion} concludes the paper.

\section{System Model}
\begin{figure}[!t]
	\centering
	\includegraphics[width=0.98\columnwidth]{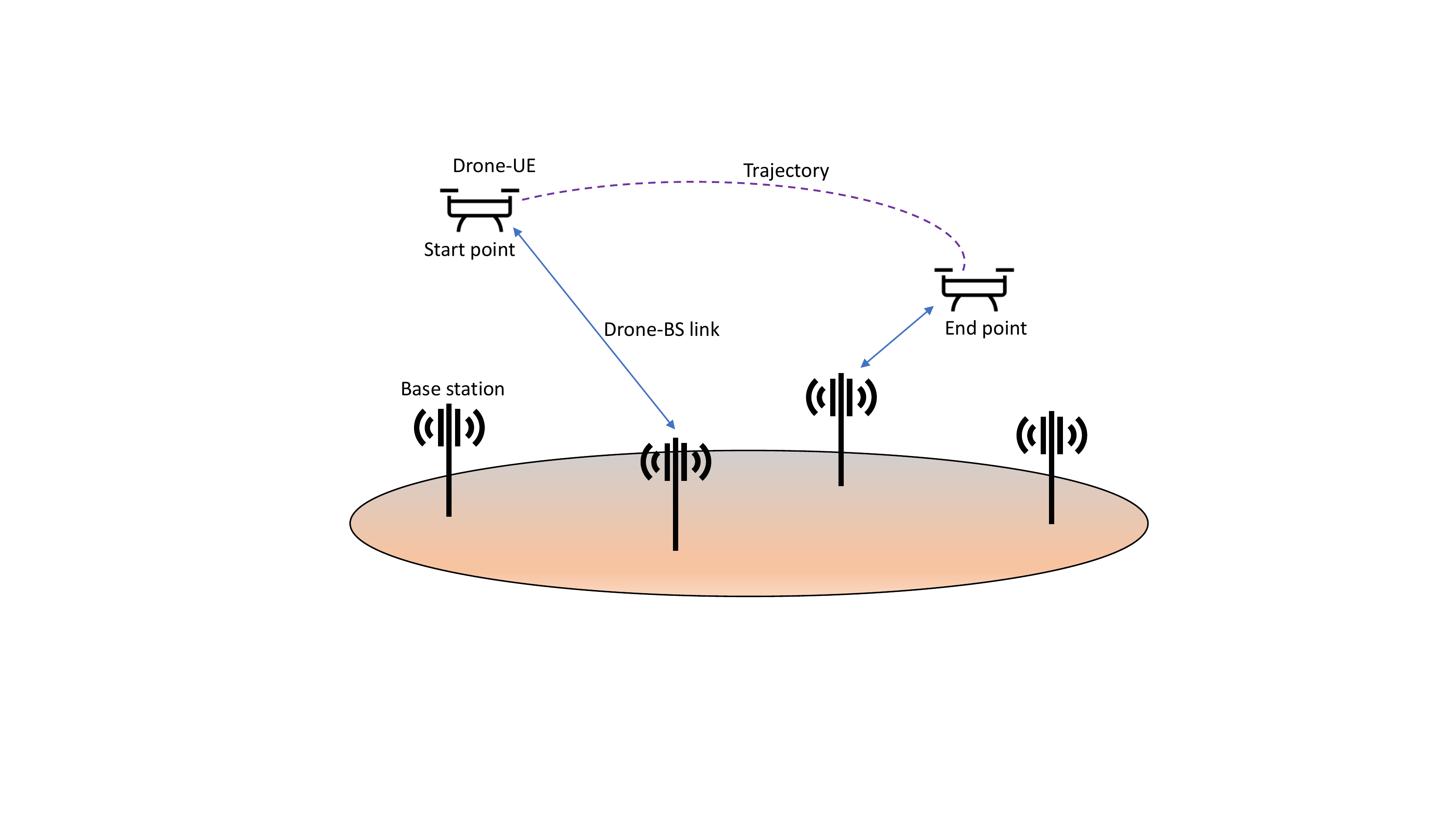}
	\caption{Illustration of the network model.} 
	\label{system}
\end{figure}

\begin{figure}[!t]
	\centering
	\includegraphics[width=\columnwidth]{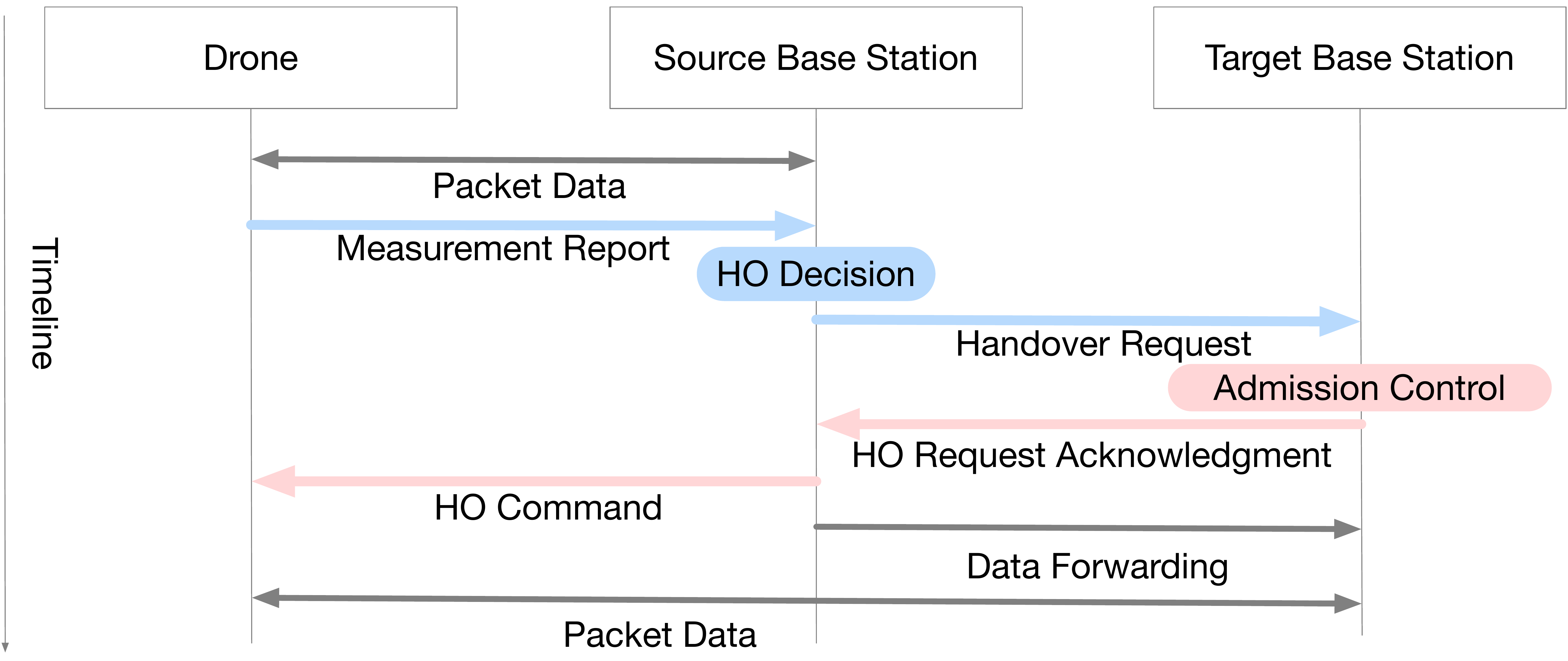}
	\caption{Illustration of a HO process.}
	\label{HO_process}
\end{figure}

We consider the scenario illustrated in Fig. \ref{system} where drone-UEs are served by a terrestrial cellular network consisting of $K$ ground BSs. We assume that a drone-UE moves along a two-dimensional (2D) trajectory at a fixed altitude which is known to the network.
To maintain reliable connectivity, the drone may perform one or more HOs during flight which changes BS-drone association. Therefore, the drone may connect to different BSs along its route. We consider predefined locations along the drone trajectory where it can perform a handover. At each such location, the drone decides: 1) whether to do a HO, and 2) the new serving BS in case a HO is needed.
As illustrated in  Fig. \ref{HO_process}, HO process typically involves several steps and signaling between drone and BSs such as measurement report, HO commands and admission control \cite{sivanesan2015mobility}. Several factors govern the outcome of a HO process such as BS distribution, received signal strength characteristics, drone speed and flight trajectory. 

In general, always connecting to the strongest BS (i.e., that provides maximum RSRP) may be detrimental for drone connectivity and HO signaling overhead. On the one hand, HO decision solely based on the current maximum RSRP can trigger many subsequent HOs during drone flight, which is not efficient. On the other hand, it can cause ping-pong HOs and connectivity failures as the signal strength fluctuates rapidly during a drone flight \cite{sivanesan2015mobility}. This motivates the need of an efficient HO mechanism which accounts for the mobility challenges facing a drone-UE in a terrestrial cellular network. Let us use RSRP as a proxy for reliable connectivity and the number of HOs as a measure of the HO signalling overhead. 
Intuitively, a desirable HO mechanism will maintain a sufficiently large RSRP while incurring a small number of HOs during a flight.   

In this paper, we propose a RL-based framework 
to determine the optimal sequential HO decisions for a drone-UE to enable reliable connectivity while accounting for the HO overhead. To this end, in our proposed RL-based HO framework, we consider two key factors in the objective function: 1) serving cell RSRP values, and 2) cost (or penalty) for performing a HO. From a design perspective, it is desirable to strike a balance between maximizing the RSRP values and minimizing the number of HOs. Furthermore, to flexibly adjust the impact of the number of HOs and serving cell RSRP values in the HO decisions, we consider $w_{HO}$ and $w_{RSRP}$ as the weights of number of HOs and serving cell RSRP.

\section{Background of RL}\label{Background of RL}
\begin{figure}[!t]
	\centering
	\includegraphics[width=.7\columnwidth]{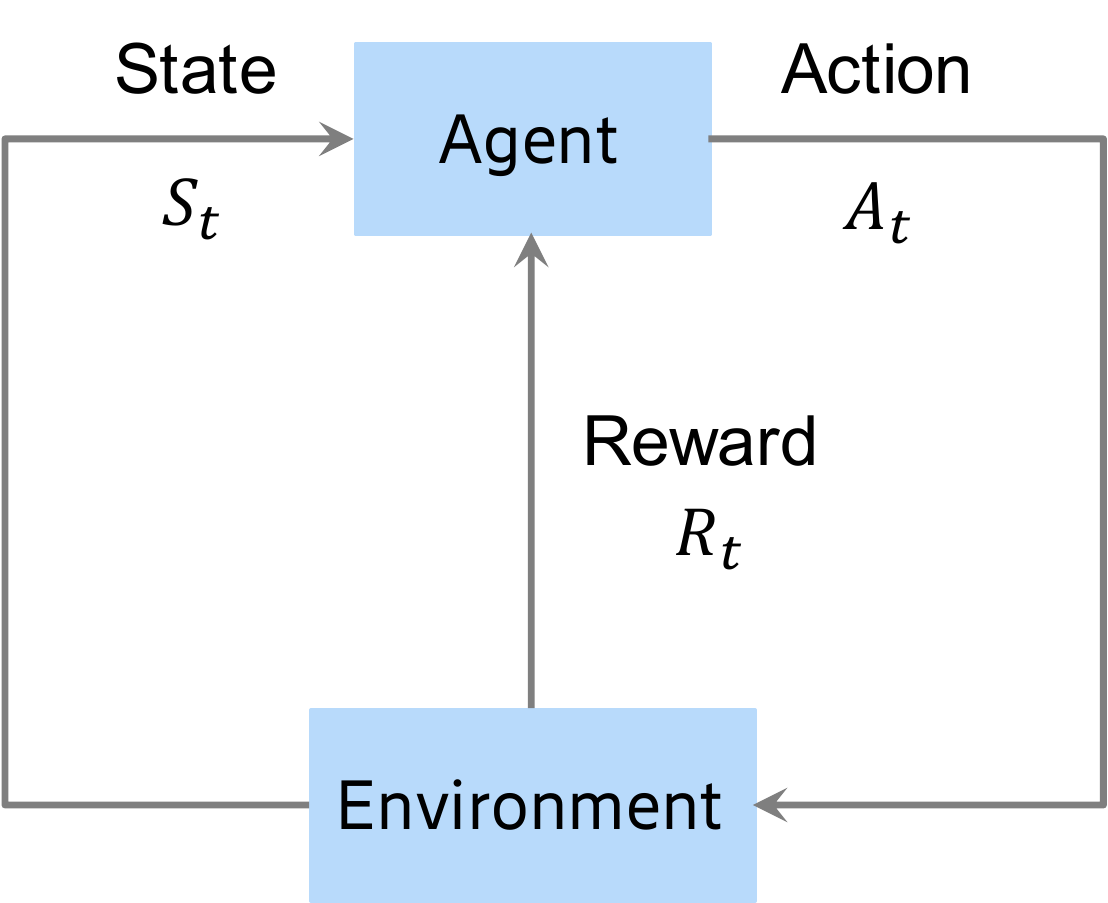}
	\caption{Illustration of RL \cite{sutton1998introduction}.}
	\label{RL}
\end{figure}

RL is a learning algorithm where an agent interacts with an environment by taking actions based on the current state and the anticipated future rewards\cite{sutton1998introduction}. As illustrated in Fig. \ref{RL}, the agent observes state $S_t$ and takes action $A_t$ at time $t$. It receives feedback in the form of a reward $R_t$ and chooses subsequent actions to maximize the expected reward accumulated over time. %
RL is often described using a Markov decision process characterized by a tuple $(\mathcal{S},\mathcal{A},\{{P}_{sa}\},\lambda,{R})$, where $\mathcal{S}$ is a set of states, $\mathcal{A}$ denotes the set of actions, ${P}_{sa}$ gives the state transition probabilities for a state $s\in\mathcal{S}$ and action $a\in\mathcal{A}$, $\lambda\in[0,1)$ gives the discount factor, and $R:\mathcal{S}\times\mathcal{A}\rightarrow\mathbb{R}$ denotes the reward function. With this information, the Markov decision process can be solved to get the optimal policy, i.e., the action to take at each state such that the expected sum of discounted rewards is maximized.

Q-learning \cite{Q-learning} is a type of model-free RL where the goal is to learn the optimal policy for the given Markov process in the absence of $\{{P}_{sa}\}$ and $R$. 
Let us define the Q-value $Q^{\pi}(s,a)$ for a policy $\pi$ as the expected sum of discounted rewards when the agent takes an action $a$ in state $s$ and chooses actions according to the policy $\pi$ thereafter. Using an iterative process, the agent will eventually learn the optimal Q-values $Q^*(s,a)$ over time. The actions with the highest Q-values for each state constitute the optimal policy \cite{Q-learning,sutton1998introduction}. 
With a slight abuse of notation, we use $Q_{t}(s,a)$ to denote the Q-value at time $t$ during the iterative process. When the agent performs an action $a$ in a state $s$ at time $t$, it receives an immediate reward $R_{t+1}$ and transitions to state $s'$. The new Q-value can be evaluated using
\begin{align}
\label{Q_iteration}
Q_{t+1}(s,a)\leftarrow (1-\alpha)~ Q_t(s,a)+\alpha\left[R_{t+1}+\lambda \max \limits_{a'\in\mathcal{A}}Q_t(s',a')\right]
\end{align}
where $\alpha$ is the learning rate. With this approach, Q-learning computes the optimal values for all states at once using successive approximations \cite{Q-learning,sutton1998introduction}.

\begin{algorithm}[!t]  
	\caption{Q-value iteration for drone HO scheme using Q-learning}  
	\label{algorithm_Q}  
	{\small
		\begin{algorithmic}[1]   
			\State Initialize input parameters: 
			\Statex Drone trajectory $\mathcal{R}=\{\mathcal{P}_i|i=0,1,...,l-1\}$;
			\Statex Set $\textbf{Q}
			\leftarrow \mathbf{0}_{l\times k\times k}$; ${\textbf{HO}}_{s,s'}\leftarrow \mathbf{0}_{ k\times k}$;
			\Statex Set $w_{HO}$, $w_{RSRP}$, $\epsilon$, $\alpha$, $\lambda$; \Statex$C_{k_{s}}\leftarrow k$ strongest cells at starting waypoint $\mathcal{P}_0$;
			\For{$i$ in length($\mathcal{R}$)$-1$}
			\State $C_{k_{s'}}\leftarrow k$ strongest cells at  waypoint $\mathcal{P}_{i+1}$;
			\State $RSRP_{k_{s'}}\leftarrow$ RSRP values for cells in $C_{k_{s'}}$;
			\State $\forall~ y\in\mathcal{I}_k$, set $\textbf{Q}[i,:,y]\leftarrow w_{RSRP}\cdot RSRP_{k_{s'}}$;
			\State Binary matrix ${\textbf{HO}}_{s,s'}\leftarrow C_{k_{s}}\neq C_{k_{s'}}$;
 			\State $\textbf{Q}[i,:,:]\leftarrow \textbf{Q}[i,:,:]-w_{HO}\times {\textbf{HO}}_{s,s'}$;
     		\State $C_{k_{s}}\leftarrow C_{k_{s'}}$;
			\EndFor
			\State Reward matrix $\textbf{R} \leftarrow  \textbf{Q}$; 
			\While{ training step $<$ $n$}
			\State $j = 0$;
			\Statex \hskip\algorithmicindent$\epsilon$-greedy algorithm:
			\For{$i$ in length($\mathcal{R}$)$-1$}
			\If{$\epsilon>$ uniform random value on interval [0,1] }
			\State 	{$j_\text{new}$} $\leftarrow \argmax\limits_{u\in\mathcal{I}_k}\textbf{Q}[i, j, u]$;
			\Else
			\State $j_\text{new}\leftarrow$ pick a random number from $\mathcal{I}_k$;
			\EndIf
			\Statex \hskip\algorithmicindent Update Q-values:
			\State $Q=\textbf{Q}[i, j, j_{\text{new}}]$;
			\State $Q=( 1 - \alpha )\cdot Q+
			\alpha\cdot \textbf{R}[i, j, j_{\text{new}}]$ \par 
			\hskip\algorithmicindent$\qquad+~
			\alpha \lambda\argmax\limits_{v\in\mathcal{I}_k}\textbf{Q}[i + 1, j_{\text{new}}, v]$;
			\State $\textbf{Q}[i, j, j_{\text{new}}]=Q$;
			\EndFor
			\State $j=j_{\text{new}}$;
			\EndWhile
			\State \Return $\textbf{Q}$
	\end{algorithmic}
}
\end{algorithm}

 \section{RL-Based HO Optimization Framework}\label{Model and Algorithms}
In this section, we formally define the state, action and reward for the considered scenario. The objective is to determine the HO decisions for each waypoint along the given route. We also propose an algorithm based on Q-learning to obtain optimal HO decisions for the given route. In Table \ref{DefinitionsModel}, we list the main parameters used in the proposed RL-based HO optimization framework.
\begin{table}[!t]
 	\caption{Definitions in our model related to RL.}
\resizebox{\columnwidth}{!}{
\begin{tabular}{@{}ll@{}}
\toprule
Label & Definition \\ \midrule
 $\mathds{I}(HO)$    &       HO cost   \\
 $w_{HO}$    &       Weight for HO cost   \\
 $w_{RSRP}$    &       Weight for serving cell RSRP   \\
  $R$    &       Reward defined as the weighted combination of HO cost and RSRP  \\
  $S$ &        State defined as $[x_s,y_s,\theta_s,c_s]$     \\   
   $(x_s, y_s)$ &       Position coordinate at state $S=s$    \\ 
    $\theta_s$ &        Movement direction at state $S=s$     \\ 
     $c_s$ &        Serving cell at state $S=s$     \\ 
   $S'$ &        Next state of $S$     \\ 
   $A$ &       Action performed at state $S$     \\    
   $A'$ &       Action performed at state $S'$    \\
   $Q(s,a)$   &    Q-value of taking action $a$ at state $s$      \\
   $\alpha$    &       Learning rate     \\
   $\lambda$   &       Discount factor  \\
   $\epsilon$   &       Exploration coefficient  \\
   $n$ & Number of training episodes \\
 \bottomrule
\end{tabular}}

\label{DefinitionsModel}
\end{table}
 \subsection{Definitions}
\textbf{State:} The state of a drone represented by $S= [x_s,y_s,\theta_s,c_s]$ consists of the drone's position $\mathcal{P}_s:(x_s,y_s)$, its movement direction $\theta_s\in \{k\pi/4, k=0,...,7\}$, and the currently connected cell $c_s\in C$, where $C$ is the set of all candidate cells. 
We describe how a drone trajectory is generated in our model given an initial location $\mathcal{P}_o$ and a final location $\mathcal{P}_e$ of the drone.
 At the initial location, the movement direction resulting in the shortest path to the final location is selected. The drone moves in the selected direction for a fixed distance until it reaches the next waypoint. The same procedure is repeated at each waypoint until it reaches closest to the final location. We note that the resulting drone trajectory is not necessarily a straight line due to a finite number of possible movement directions in our model. We recall that the RL-based HO algorithm merely expects that the drone trajectory is known beforehand. That is, it is not significant how the fixed trajectories are generated but we describe the methodology for the sake of completeness. 

\begin{figure}[!t]
\centering
 \subfloat[Illustration of current and next state.]{%
\label{Model_part1}
 \includegraphics[width=\columnwidth]{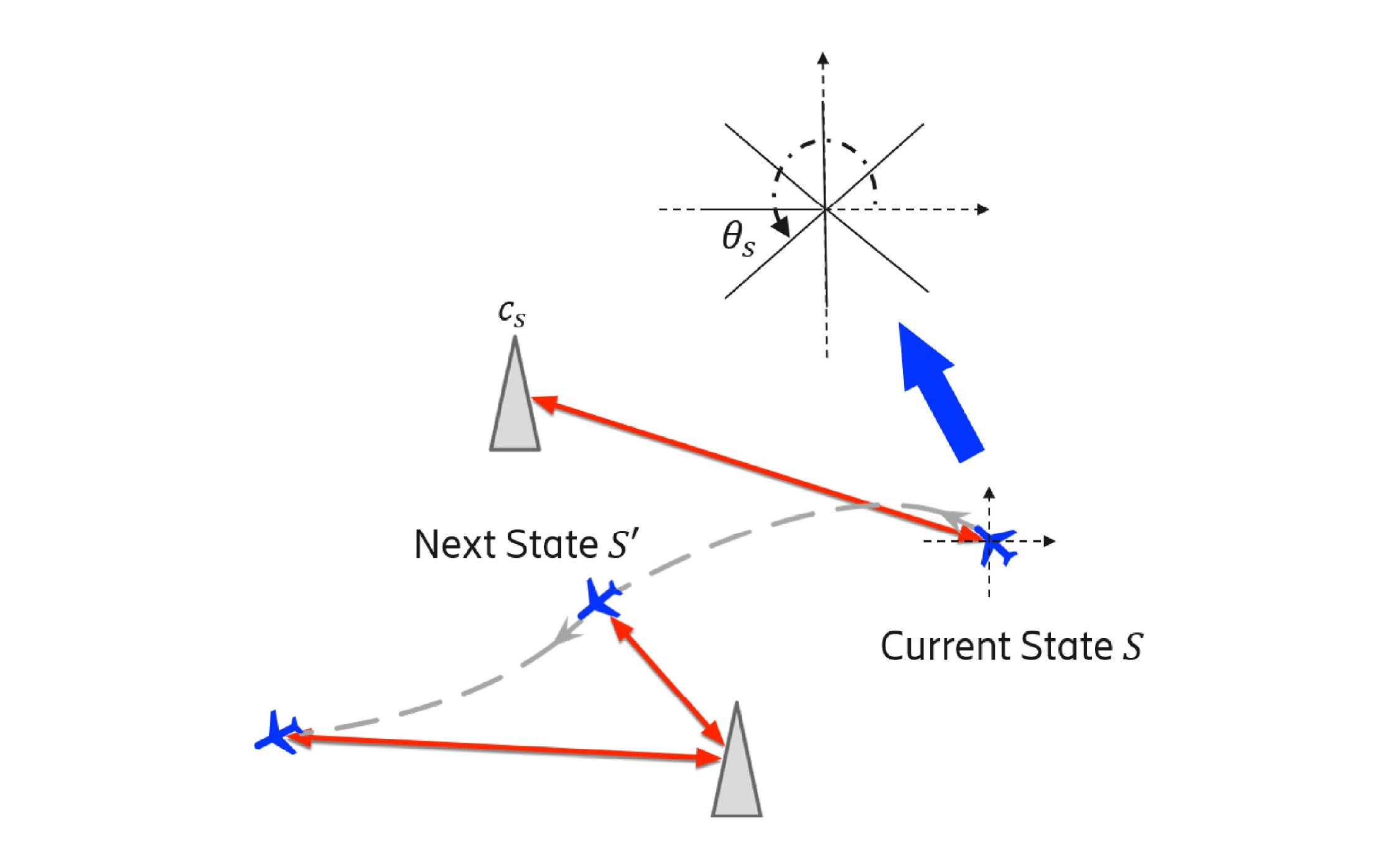}}\hfill
 \subfloat[Illustration of action.]{%
\label{Model_part2}
  \includegraphics[width=\columnwidth]{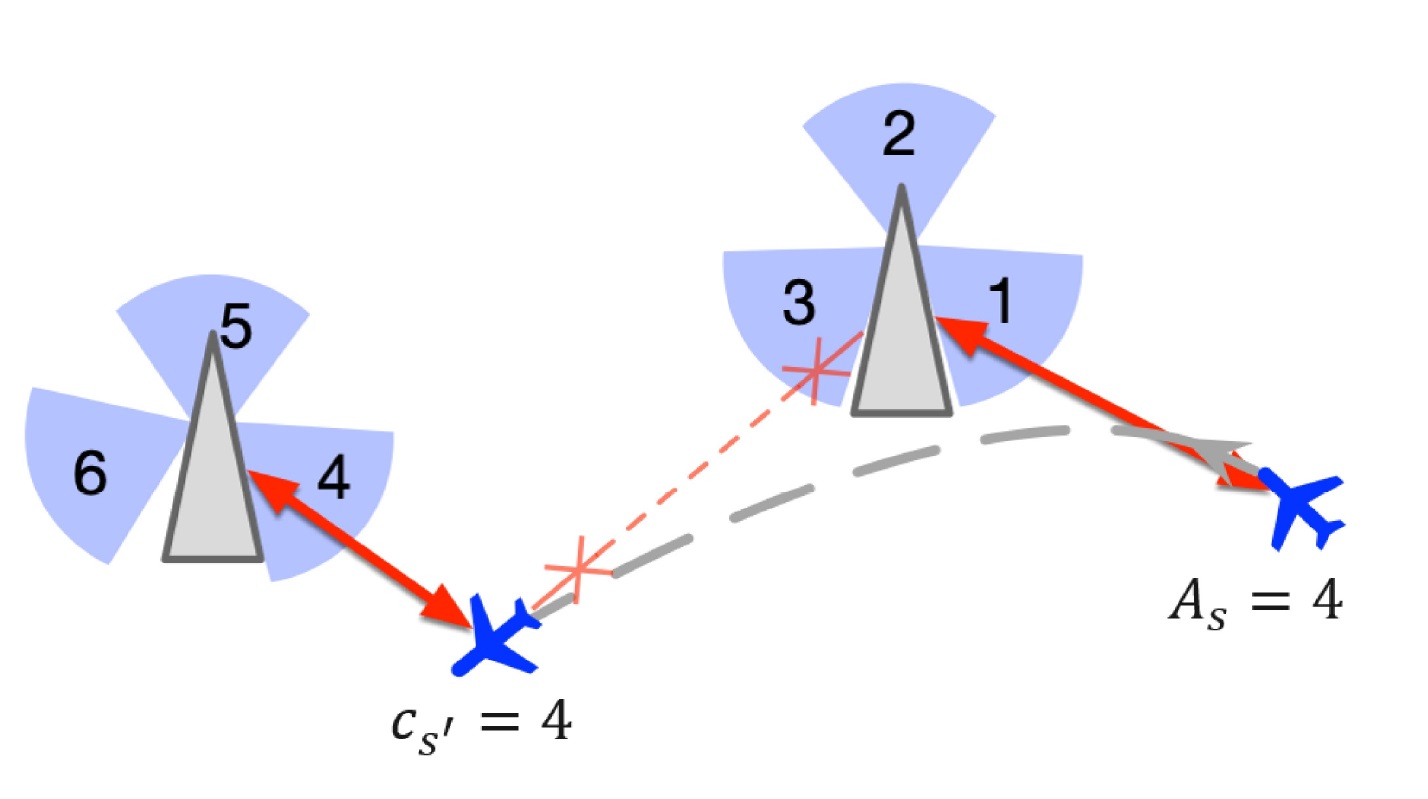}}
 
 \subfloat[An example of HO decisions during a trip.]{%
\label{Model}
 \includegraphics[width=.98\columnwidth]{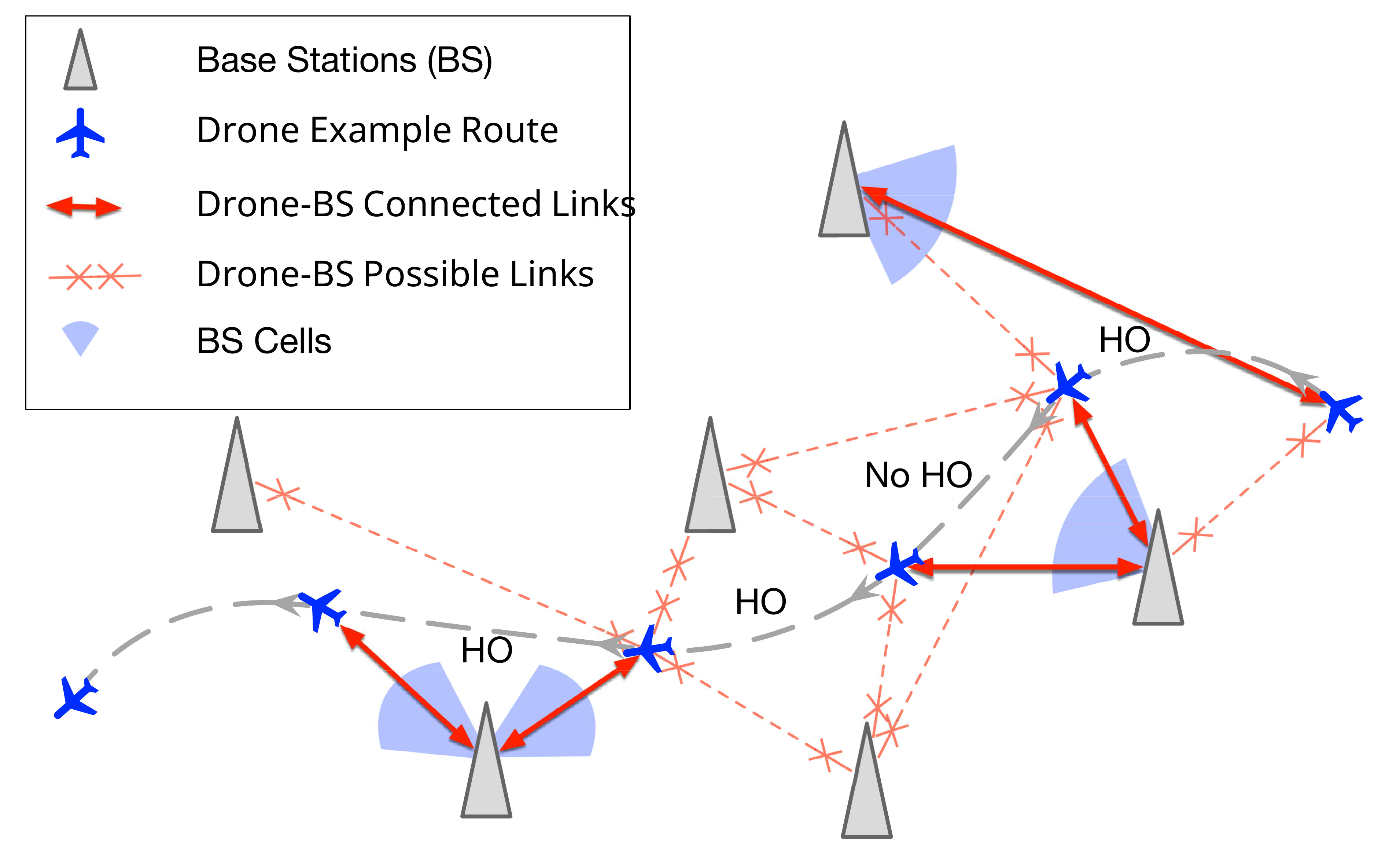}}\hfill 
  \caption{Illustration of the proposed RL-based framework.}
\label{Model_parts}
\end{figure}

\textbf{Action:} The drone's action $A_s$ at current state $s$ corresponds to choosing a serving cell  for the next state $s'$. For example, as shown in Fig. \ref{Model_part2}, if $A_s=4$, then at state $s'$, the drone switches to the cell $c_{s'}=4$.

\textbf{Reward:} We now define a reward function to encourage the desired HO behavior. 
As shown in Fig. \ref{Model}, the serving cells need to be decided along the trajectory and the
goal is to reduce the number of HOs as well as maintain reliable connectivity. 
During a flight, the drone need not only focus on the signal strength at the current location. Rather, it might as well connect to a cell with a lower RSRP at one waypoint that results in fewer HOs at the subsequent waypoints. To achieve a balance between the two conflicting goals, our model considers a weighted combination of the HO cost and the serving cell RSRP at future state as the reward function
\begin{align}
    R=-w_{HO}\times \mathds{I}(HO)+w_{RSRP}\times RSRP_{s'},
\end{align}
where $w_{HO}$  and $w_{RSRP}$ respectively denote  the weights for the HO cost and the RSRP, while $\mathds{I}(HO)$ is the indicator function for HO, i.e., $\mathds{I}(HO)=1$ when the serving cells at states $s$ and $s'$ are different and $\mathds{I}(HO)=0$ otherwise.
\subsection{Algorithm of HO Scheme using Q-learning}
For complexity reduction, the action space $\mathcal{A}$ in our model is restricted to the strongest $k$ candidate cells for every state. Let us define a set $\mathcal{I}_k=\{0,1,\cdots,k-1\}$. For a trajectory with $l$ waypoints, 
the resulting Q-table $\textbf{Q}\in \mathds{R}^{l\times k\times k}$ is stored in memory and updated according to (\ref{Q_iteration}). The stepwise iterative process is given in Algorithm \ref{algorithm_Q}. The algorithm complexity is ${\rm O}(cn)$, where $n$ is the total number of training episodes and the constant $c$ is given by the route length $l$. The initial Q-table for the given trajectory is generated in steps 2-9. In step 6, a binary square matrix of size $k$ is generated such that its $(p,q)$-th entry is 0 if the $p$-th strongest cell at state $s$ is the same as the $q$-th strongest cell in state $s'$, and it is 1 otherwise.    
The Q-value iterations for each training episode are performed in step 11-24, where steps 14-18 perform the $\epsilon$-greedy exploration \cite{sutton1998introduction} while step 20 implements (\ref{Q_iteration}). Finally, values for choosing different actions are stored in $\textbf{Q}$ where the highest value represents the optimal choice. Hence, a sequence of HO decisions for the waypoints of the given route can be obtained according to the maximal Q-value at each state.

\section{Simulation Results}\label{Experiments and Results}
In this section, we evaluate the performance of the proposed RL-based HO mechanism.
For performance comparison, we consider a greedy HO scheme as the baseline in which the drone always connects to the strongest cell. For each flight trajectory, we calculate a performance metric called \emph{HO ratio} which we define as the ratio of the number of HOs using the proposed scheme to that for the baseline scheme. Thus, the HO ratio is always $1$ for the baseline case. To depict the tradeoff between the number of HOs and the observed RSRP values, we evaluate the performance for different weight combinations of $w_{HO}$ and $w_{RSRP}$ in the reward function. 
For the special case when there is no HO cost, the proposed RL-based HO scheme is equivalent to the baseline.  As the ratio $\frac{w_{HO}}{w_{RSRP}}$ increases, the number of HOs decreases and the HO ratio approaches zero. 
\subsection{Data Pre-processing}
\begin{figure}[!t]
	\centering
		
		\includegraphics[width=1\columnwidth]{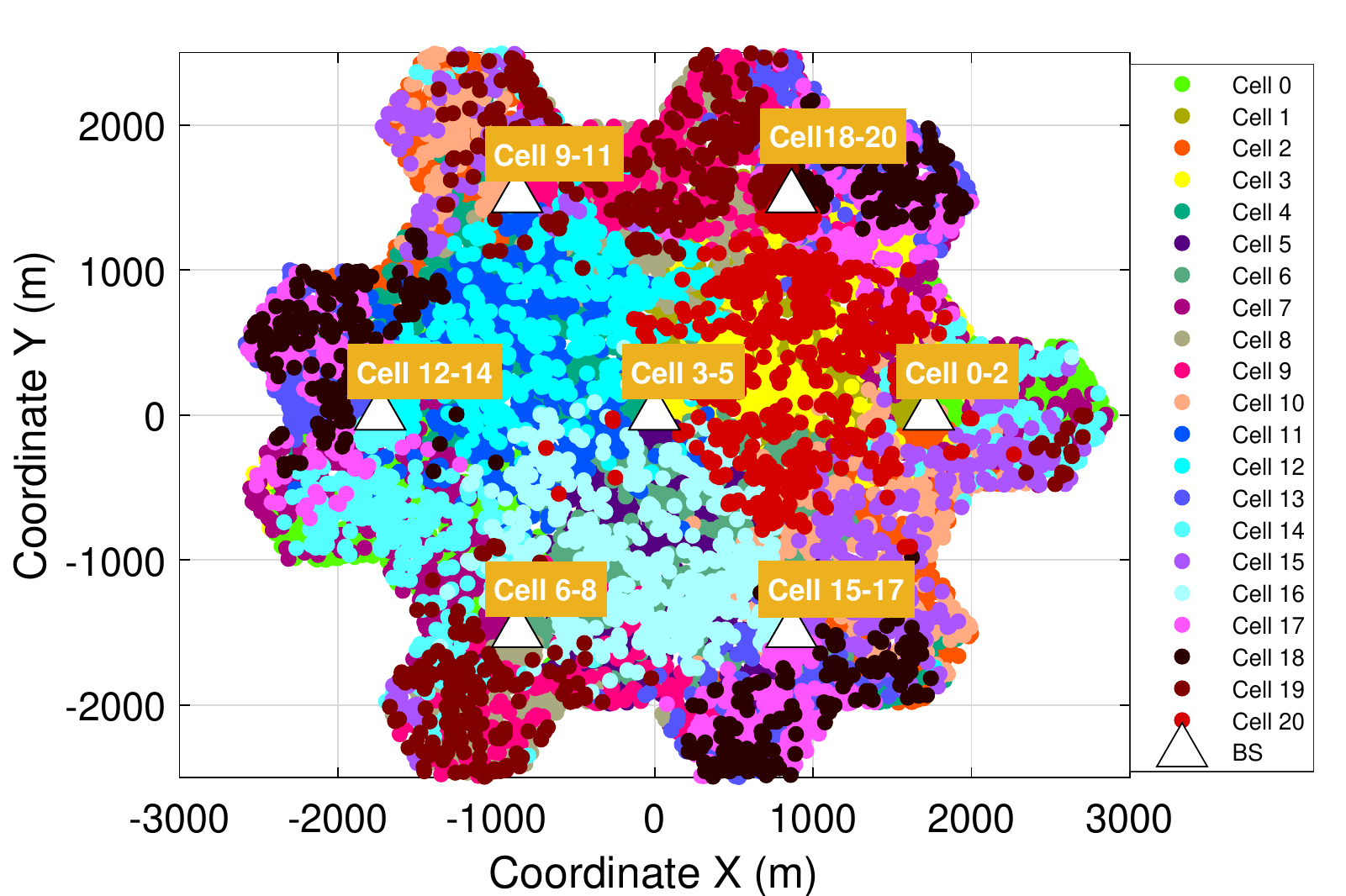}
		\caption{Strongest cell association map of RSRP samples without quantization.}
		\label{samples}
\end{figure}

\begin{figure}
\centering	
		\includegraphics[width=1\columnwidth]{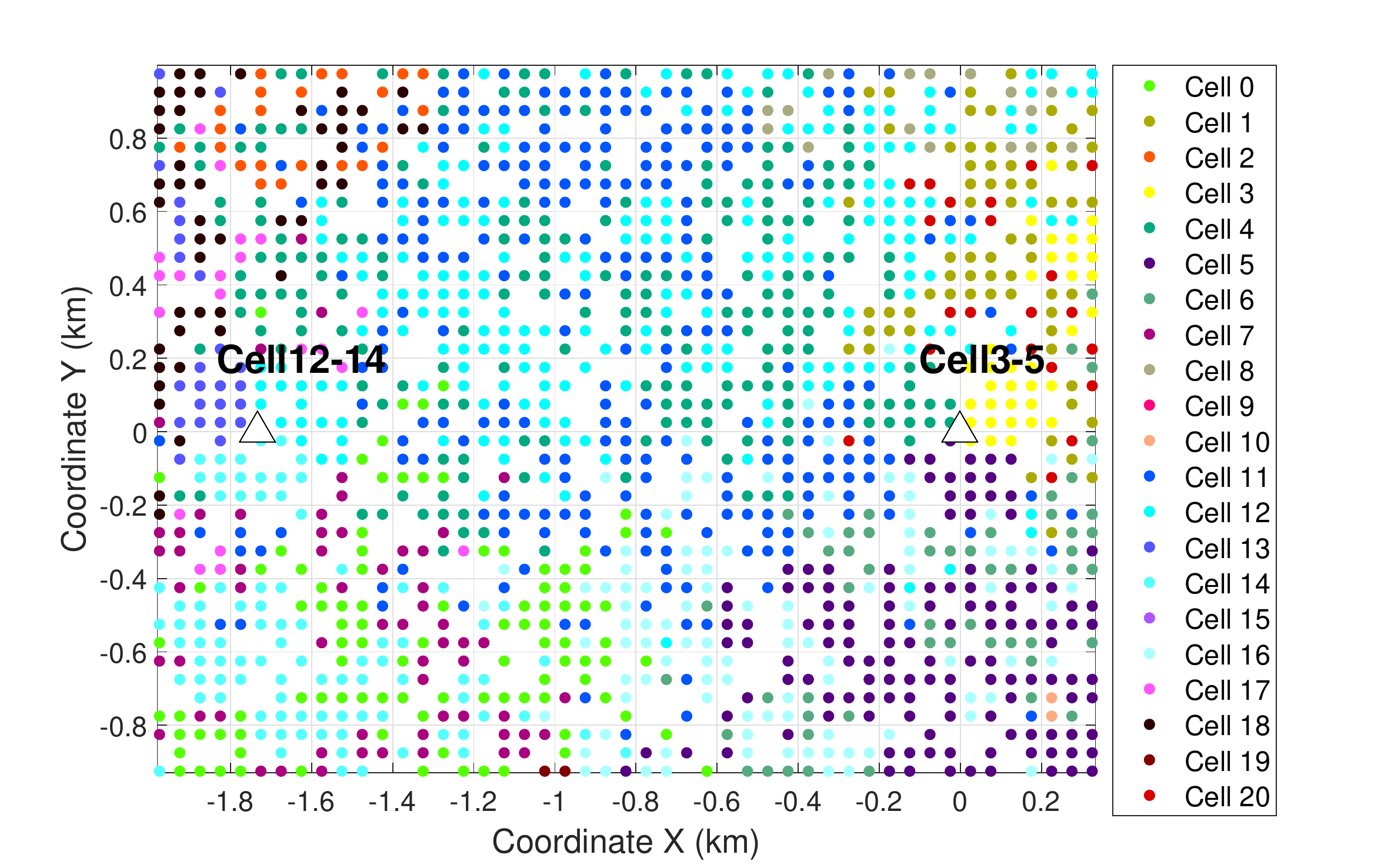}	
	\caption{Cell association map based on collected RSRP samples after quantization.}
		\label{quansamples}
\end{figure}

In our simulations, we consider a deployment of $7$ BSs in a 2D geographical area of $6\times 5$ km$^2$ where each BS has $3$ cells or sectors. We collect $10000$ samples of RSRP values corresponding to each of these $21$ cells for different UE locations at an altitude of $50$ m, as shown in Fig. \ref{samples}. For normalization, the RSRP samples thus obtained are linearly transformed  to the interval [0 1].
To further quantize the considered 
space, as shown in Fig. \ref{quansamples}, we partition the area into bins of size $50\times 50$ m$^2$. For each bin, we calculate the represntative RSRP value for a cell as the average of the RSRP samples in that bin. 

\subsection{Results using Q-learning}\label{Results using Q-learning}

\begin{figure}[!t]
	\centering
	\includegraphics[width=.98\columnwidth]{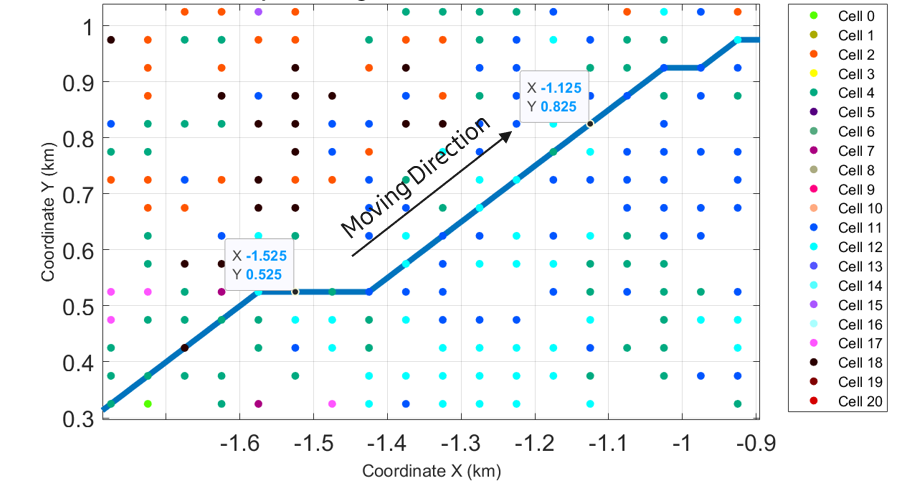}
	\caption{A zoomed snapshot of a drone trajectory along with the serving cells.}
	\label{temp_traj}
\end{figure}

We simulate the performance using $2000$ runs for the proposed and baseline schemes. 
For each run, the testing route is generated randomly as explained in Section \ref{Model and Algorithms}. As an illustrative example, Fig. \ref{temp_traj} shows a portion of the drone trajectory along with the strongest cell at each waypoint.
In our simulations, we set $n=1000$, $\lambda=0.3$, $\alpha=0.5$, and $\epsilon=0.2$.

\begin{figure}[!t]
\centering
{%

 \includegraphics[width=1.05\columnwidth]{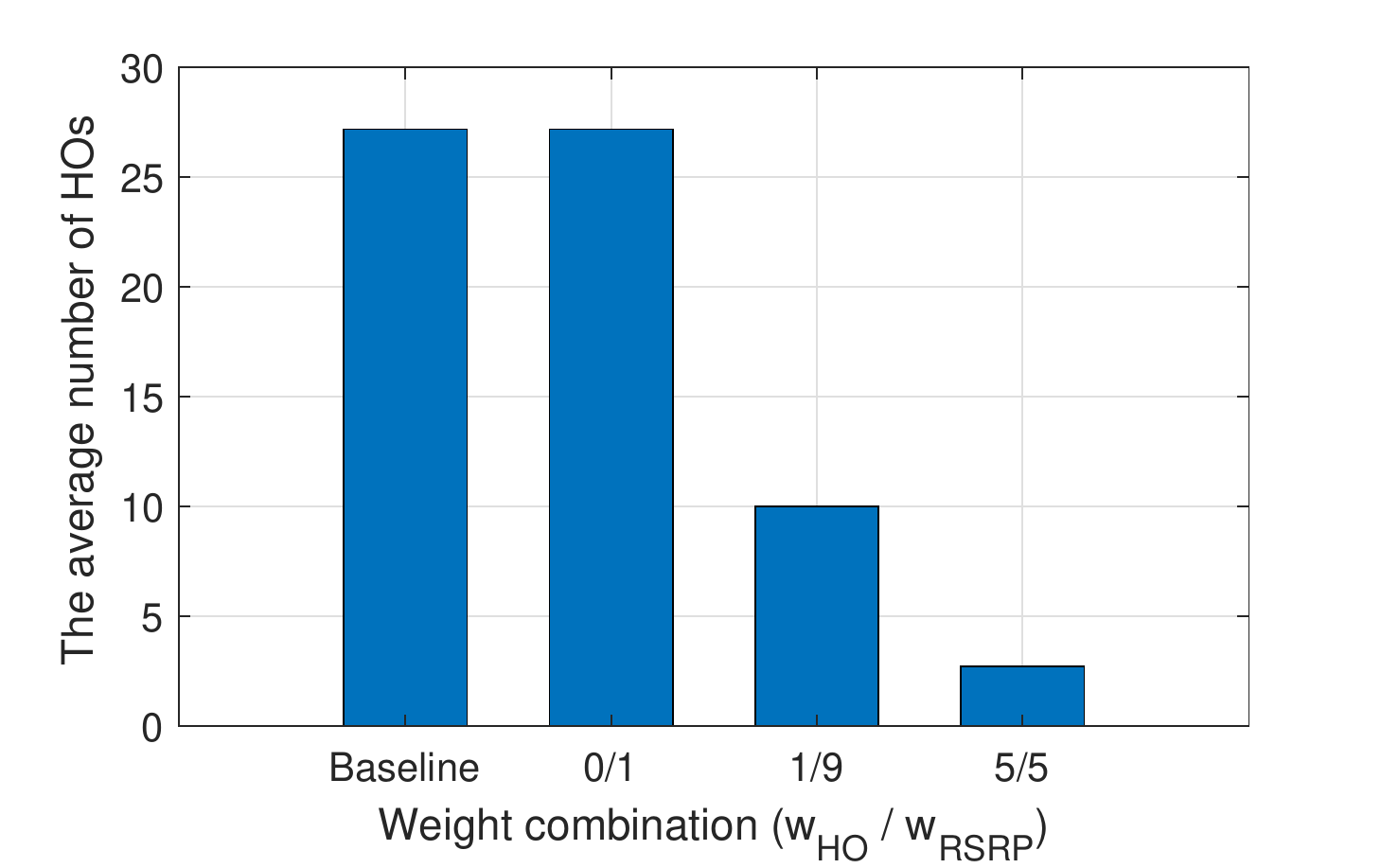}}
\caption{Average number of HOs for various weight combinations ($w_{HO}$, $w_{RSRP}$).}\label{ab_Q}
\end{figure}
\begin{figure}[!t]
\centering
{%

  \includegraphics[width=1.04\columnwidth]{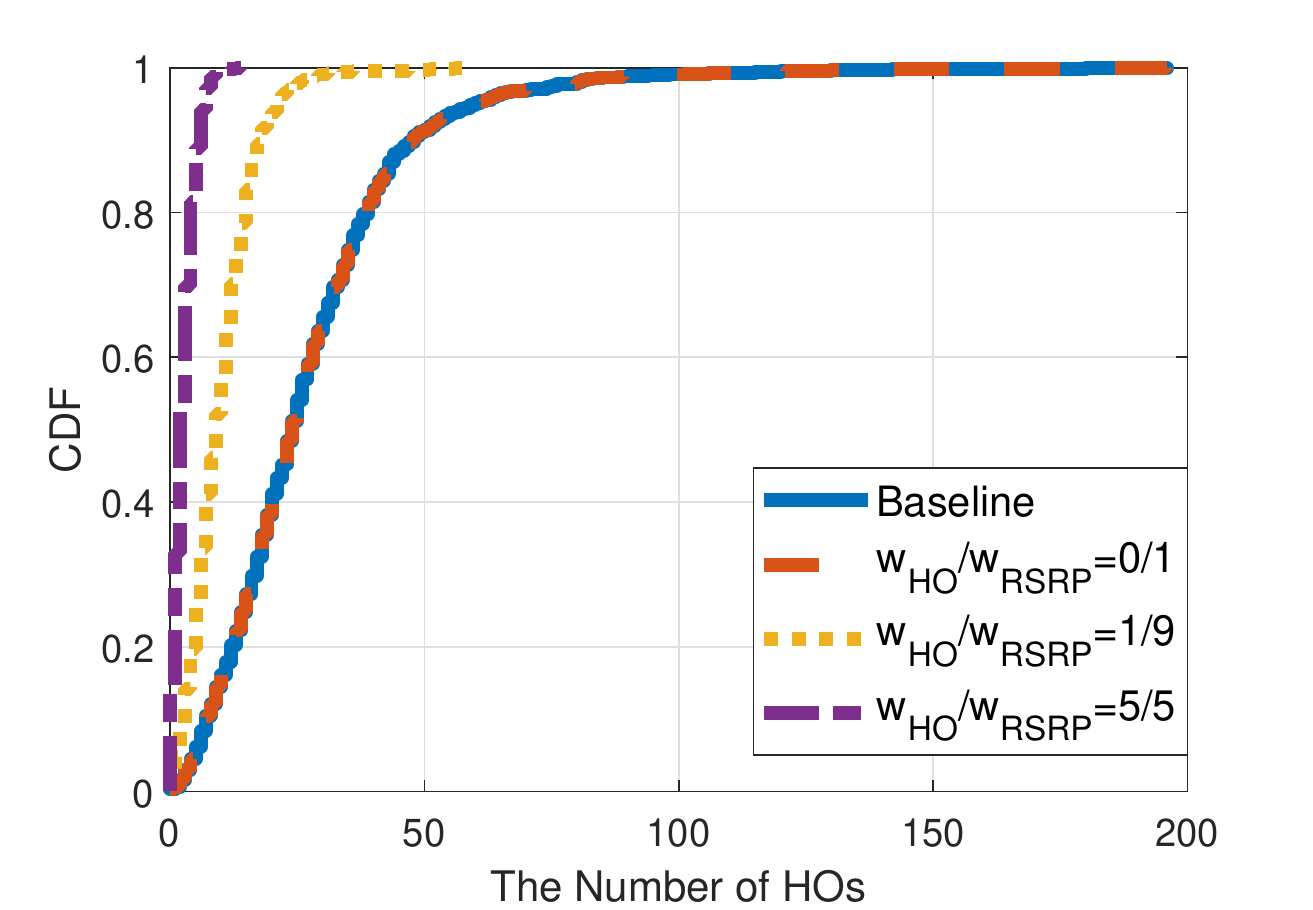}}
\caption{CDF of the number of HOs.}\label{CDF_HO_Q}
\end{figure}

\begin{figure}[!t]
	\centering
	{%
		
		\includegraphics[width=1.06\columnwidth]{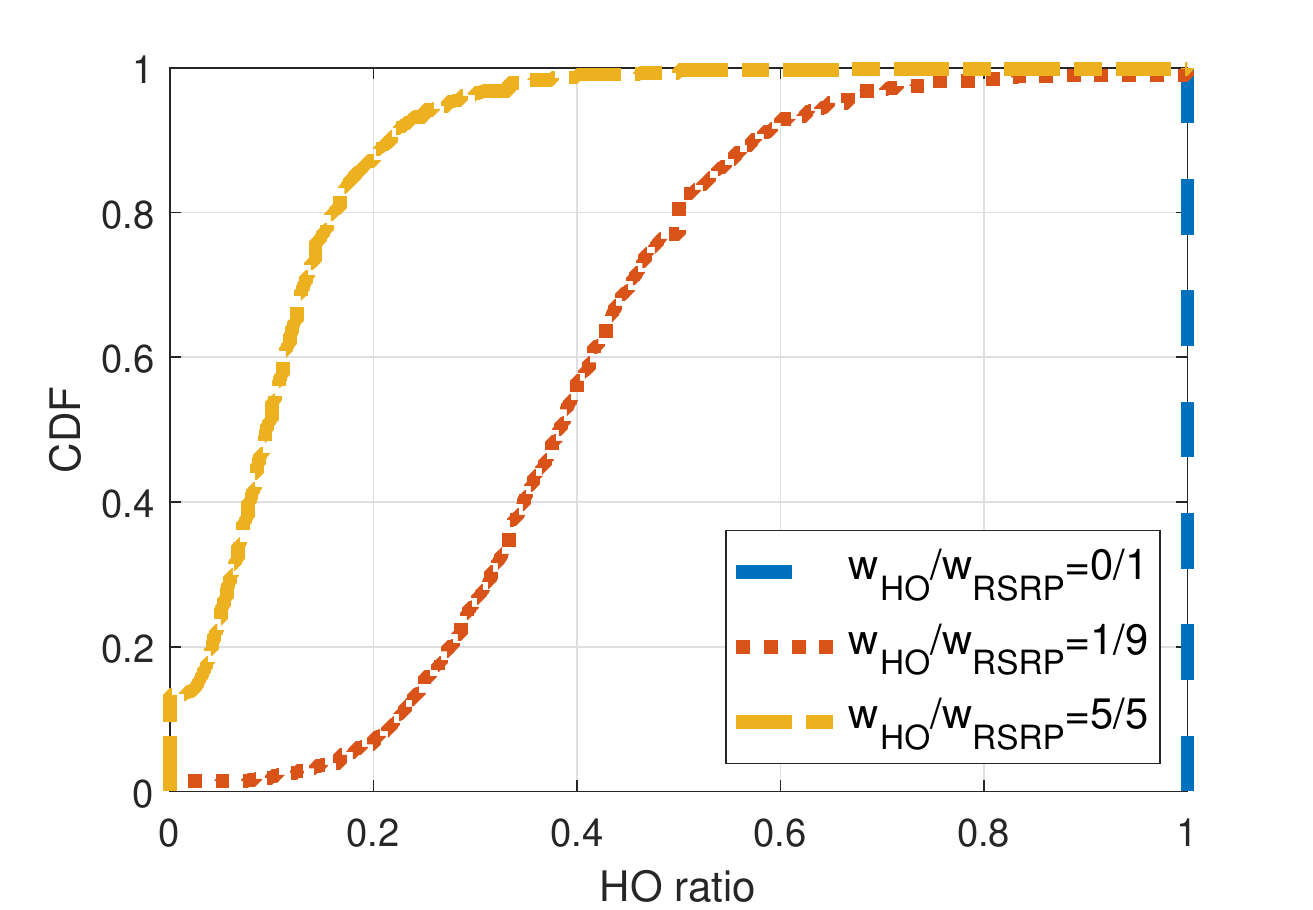}}
	\caption{CDF of the HO ratio.} \label{CDF_HOratio_Q}
\end{figure}

\begin{figure}[!t]
\centering
\includegraphics[width=1.06\columnwidth]{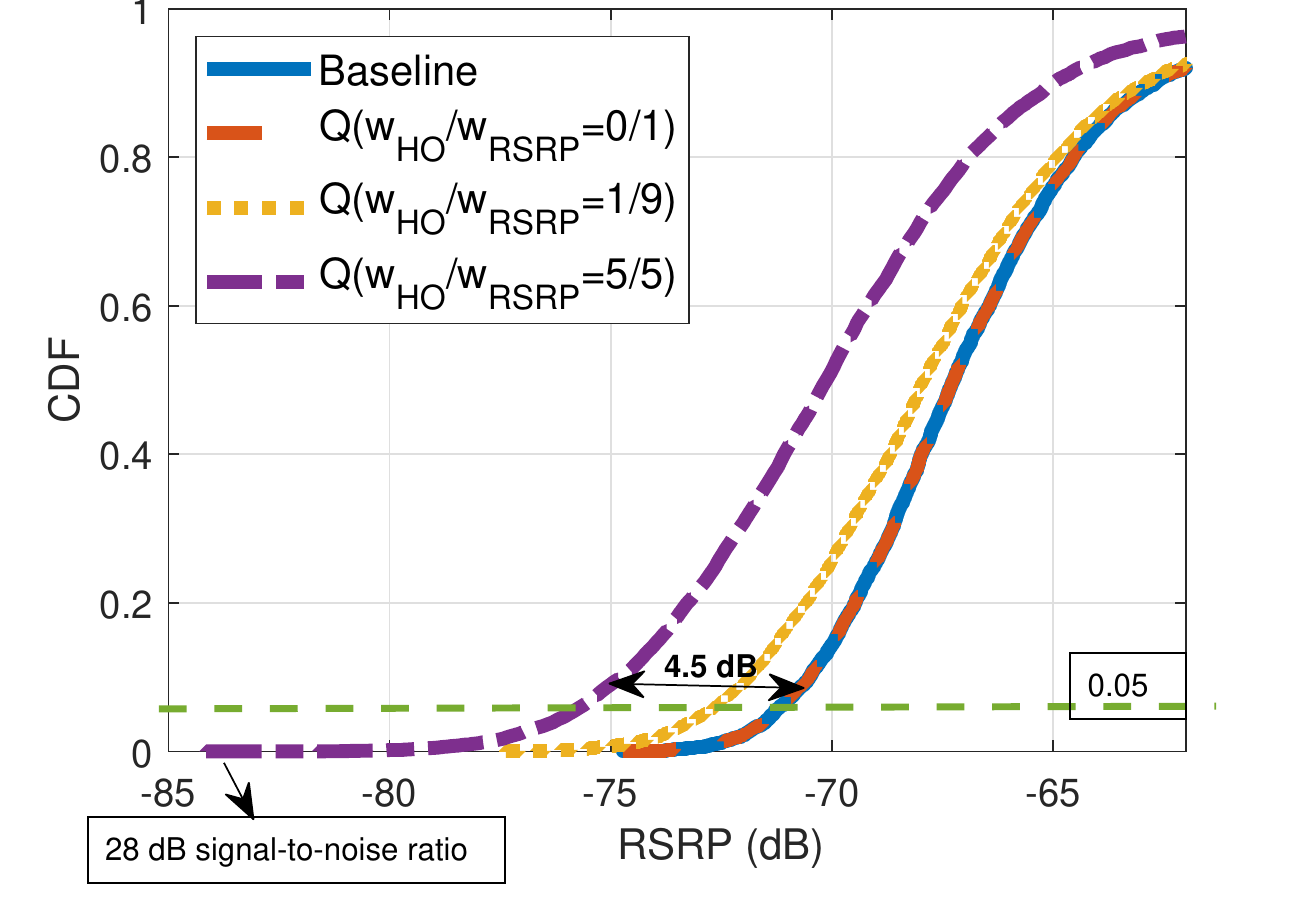} \vspace{-0.1cm}
\caption{CDF of RSRP (dBm) for various weight combinations.}
\label{CDF_RSRP_Q}
\end{figure}

 In Fig. \ref{ab_Q}, we plot the average number of per-flight HOs for different weight combinations.  The proposed scheme is equivalent to the baseline when there is no HO cost. By increasing $\frac{w_{HO}}{w_{RSRP}}$, the cost of HO increases and our approach avoids unnecessary HOs. For instance, compared to the baseline, the RL-based HO scheme can reduce the number of HOs by $82.9\%$ when $\frac{w_{HO}}{w_{RSRP}}=1$. 

In Fig. \ref{CDF_HO_Q}, we plot the cumulative distribution function (CDF) of the number of HOs in a flight. For the special case $w_{HO}=0$, we observe that the CDF for the proposed approach coincides with that of the baseline. Moreover, by properly adjusting the weights for the HO cost and RSRP, the RL-based scheme can significantly reduce the number of HOs. Similar trends can be observed for the HO ratio in Fig. \ref{CDF_HOratio_Q}. For example, for $\frac{w_{HO}}{w_{RSRP}}=\frac{1}{9}$, the number of HOs can be reduced by at least 50\% with a probability 0.8.   


In Fig. \ref{CDF_RSRP_Q}, we plot the CDF of the RSRP seen by the drone-UE for various HO costs. As expected, the RL-based HO scheme is equivalent to the baseline in terms of the RSRP distribution when there is no cost associated with a HO. This is because for the baseline case, the drone always connects to the cell offering the largest RSPR during its flight. As noted previously, the proposed RL-based scheme is flexible in that it allows reducing the ping-pong HOs (and resulting signalling overheads) at the expense of a lower RSRP. For example, when $\frac{w_{HO}}{w_{RSRP}}=1$, a (worst-case) 5th-percentile UE incurs an RSRP loss of around 4.5 dB relative to the baseline. Setting $\frac{w_{HO}}{w_{RSRP}}=\frac{1}{9}$ suffers from only a small loss in RSRP.
As evident from Fig. \ref{CDF_RSRP_Q} and Fig. \ref{CDF_HO_Q}, both choices significantly reduce the number of HOs compared to the baseline. Such a tradeoff may still be acceptable depending on the operating conditions. For instance, the minimum serving cell RSRP in our results is always greater than -85 dBm (or 28 dB signal-to-noise ratio (SNR) assuming a bandwidth of 1 MHz and a noise power of -113 dBm), which is typically sufficient to provide reliable connectivity. 
Depending on the specific scenario, the network may configure the parameters accordingly to operate at an acceptable RSRP but with a reduced HO overhead.

\section{Conclusions}\label{Conclusion}
In this work, we have proposed an RL-based HO mechanism to achieve robust drone connectivity in a cellular-connected drone network. 
Leveraging a Q-learning framework, we have provided a flexible way for HO decision making for a given flight trajectory.
We have shown how the network can trade-off the number of HOs with the received signal strength by adjusting the respective weights of these quantities in the reward function.  
 The simulation results have revealed that the proposed approach can significantly reduce the number of HOs while maintaining reliable connectivity, compared to the baseline HO scheme in which the drone always connects to the strongest cell. 

There are several potential directions for future research. First, the existing framework considers drone mobility in 2D. A natural extension will be to allow for 3D drone mobility. Second, the testing area and flying routes considered in this work are rather limited. It will be worth investigating whether our findings hold for larger testing areas and/or longer flying routes with a larger pool of candidate cells. Third, the proposed model and resulting simulations are based on the RSRP metric. Another notable contribution will be to enrich the model with additional parameters. 

\footnotesize
\bibliographystyle{IEEEtran}
\bibliography{IEEEabrv,Main}

\begin{thebibliography}{10}
\providecommand{\url}[1]{#1}
\csname url@samestyle\endcsname
\providecommand{\newblock}{\relax}
\providecommand{\bibinfo}[2]{#2}
\providecommand{\BIBentrySTDinterwordspacing}{\spaceskip=0pt\relax}
\providecommand{\BIBentryALTinterwordstretchfactor}{4}
\providecommand{\BIBentryALTinterwordspacing}{\spaceskip=\fontdimen2\font plus
\BIBentryALTinterwordstretchfactor\fontdimen3\font minus
  \fontdimen4\font\relax}
\providecommand{\BIBforeignlanguage}[2]{{%
\expandafter\ifx\csname l@#1\endcsname\relax
\typeout{** WARNING: IEEEtran.bst: No hyphenation pattern has been}%
\typeout{** loaded for the language `#1'. Using the pattern for}%
\typeout{** the default language instead.}%
\else
\language=\csname l@#1\endcsname
\fi
#2}}
\providecommand{\BIBdecl}{\relax}
\BIBdecl

\bibitem{TutorialMO}
M.~{Mozaffari}, W.~{Saad}, M.~{Bennis}, Y.~{Nam}, and M.~{Debbah}, ``A tutorial
  on {UAVs} for wireless networks: Applications, challenges, and open
  problems,'' \emph{IEEE Communications Surveys Tutorials}, vol.~21, no.~3, pp.
  2334--2360, thirdquarter 2019.

\bibitem{fotouhi2019survey}
A.~Fotouhi, H.~Qiang, M.~Ding, M.~Hassan, L.~G. Giordano, A.~Garcia-Rodriguez,
  and J.~Yuan, ``Survey on {UAV} cellular communications: Practical aspects,
  standardization advancements, regulation, and security challenges,''
  \emph{IEEE Communications Surveys \& Tutorials}, 2019.

\bibitem{yang2018telecom}
G.~Yang, X.~Lin, Y.~Li, H.~Cui, M.~Xu, D.~Wu, H.~Ryd{\'e}n, and S.~B. Redhwan,
  ``A telecom perspective on the internet of drones: From {LTE}-advanced to
  {5G},'' \emph{arXiv preprint arXiv:1803.11048}, 2018.

\bibitem{Sky}
X.~{Lin}, V.~{Yajnanarayana}, S.~D. {Muruganathan}, S.~{Gao}, H.~{Asplund},
  H.~{Maattanen}, M.~{Bergstrom}, S.~{Euler}, and Y.~.~E. {Wang}, ``The sky is
  not the limit: {LTE} for unmanned aerial vehicles,'' \emph{IEEE
  Communications Magazine}, vol.~56, no.~4, pp. 204--210, April 2018.

\bibitem{MobilieDrones}
X.~{Lin}, R.~{Wiren}, S.~{Euler}, A.~{Sadam}, H.~{Maattanen},
  S.~{Muruganathan}, S.~{Gao}, Y.~.~E. {Wang}, J.~{Kauppi}, Z.~{Zou}, and
  V.~{Yajnanarayana}, ``Mobile network-connected drones: Field trials,
  simulations, and design insights,'' \emph{IEEE Vehicular Technology
  Magazine}, vol.~14, no.~3, pp. 115--125, Sep.. 2019.

\bibitem{3GPPTR}
{3GPP TR 36.777}, ``Enhanced {LTE} support for aerial vehicles,'' 2017.

\bibitem{stanczak2018mobility}
J.~Stanczak, I.~Z. Kovacs, D.~Koziol, J.~Wigard, R.~Amorim, and H.~Nguyen,
  ``Mobility challenges for unmanned aerial vehicles connected to cellular
  {LTE} networks,'' in \emph{in Proc. of IEEE 87th Vehicular Technology
  Conference (VTC Spring)}, 2018, pp. 1--5.

\bibitem{MobilityS}
{S. Euler, H. Maattanen, X. Lin, Z. Zou, M. Bergstrom, and J. Sedin},
  ``Mobility support for cellular connected unmanned aerial vehicles:
  Performance and analysis,'' \emph{” arXiv:1804.04523}, 2018.

\bibitem{yajnanarayana20195g}
V.~Yajnanarayana, H.~Ryd{\'e}n, L.~H{\'e}vizi, A.~Jauhari, and M.~Cirkic,
  ``{5G} handover using reinforcement learning,'' \emph{arXiv:1904.02572},
  2019.

\bibitem{alkhateeb2018machine}
A.~Alkhateeb, I.~Beltagy, and S.~Alex, ``Machine learning for reliable mmwave
  systems: Blockage prediction and proactive handoff,'' in \emph{in Proc. IEEE
  Global Conference on Signal and Information Processing (GlobalSIP)}, 2018,
  pp. 1055--1059.

\bibitem{sutton1998introduction}
R.~S. Sutton, A.~G. Barto \emph{et~al.}, \emph{Introduction to reinforcement
  learning}.\hskip 1em plus 0.5em minus 0.4em\relax MIT press Cambridge, 1998,
  vol.~2, no.~4.

\bibitem{sivanesan2015mobility}
K.~Sivanesan, J.~Zou, S.~Vasudevan, and S.~Palat, ``Mobility performance
  optimization for {3GPP LTE HetNets},'' 2015.

\bibitem{Q-learning}
C.~J. Watkins and P.~Dayan, ``Q-learning,'' \emph{Machine learning}, vol.~8,
  no. 3-4, pp. 279--292, 1992.

\end{thebibliography}

\end{document}